\documentclass[preprint,superscriptaddress]{revtex4-1}
\usepackage{graphicx}
\usepackage{color}
\usepackage[colorlinks=true,
            linkcolor=blue,
            urlcolor=blue,
            citecolor=blue,
            hyperfootnotes=true]{hyperref}

\begin{document}
\title{Tunable double-Weyl Fermion semimetal state in the SrSi$_2$ materials class}

\author{Bahadur Singh}
%\email{bahadursingh24@gmail.com}
\affiliation{SZU-NUS Collaborative Center and International Collaborative Laboratory of 2D Materials for Optoelectronic Science $\&$ Technology, Engineering Technology Research Center for 2D Materials Information Functional Devices and Systems of Guangdong Province, College of Optoelectronic Engineering, Shenzhen University, ShenZhen 518060, China}
\affiliation{Centre for Advanced 2D Materials and Graphene Research Centre, National University 
of Singapore, Singapore 117546}

\author{Guoqing Chang}
\affiliation{Centre for Advanced 2D Materials and Graphene Research Centre, National University 
of Singapore, Singapore 117546}
\affiliation{Department of Physics, National University of Singapore, Singapore 117542}
\affiliation{Institute of Physics, Academia Sinica, Taipei 11529, Taiwan}

\author{Tay-Rong Chang}
\affiliation{Department of Physics, National Cheng Kung University, Tainan 701, Taiwan}

\author{Shin-Ming Huang}
\affiliation {Department of Physics, National Sun Yat-sen University, Kaohsiung 80424, Taiwan}

\author{Chenliang Su}
\email{chmsuc@szu.edu.cn}
\affiliation{SZU-NUS Collaborative Center and International Collaborative Laboratory of 2D Materials for Optoelectronic Science $\&$ Technology, Engineering Technology Research Center for 2D Materials Information Functional Devices and Systems of Guangdong Province, College of Optoelectronic Engineering, Shenzhen University, ShenZhen 518060, China}

\author{Ming-Chieh Lin} 
\email{mclin@hanyang.ac.kr}
\affiliation{Multidisciplinary Computational Laboratory, Department of Electrical and Biomedical Engineering, Hanyang University, Seoul 04763, South Korea}

\author{Hsin Lin}
\email{nilnish@gmail.com}
\affiliation{Centre for Advanced 2D Materials and Graphene Research Centre, National University 
of Singapore, Singapore 117546}
\affiliation{Department of Physics, National University of Singapore, Singapore 117542}
\affiliation{Institute of Physics, Academia Sinica, Taipei 11529, Taiwan}

\author{Arun Bansil}
\affiliation{Department of Physics, Northeastern University, Boston, Massachusetts 02115, USA}

\begin{abstract}
We discuss first-principles topological electronic structure of noncentrosymmetric SrSi$_2$ materials class based on the hybrid exchange-correlation functional. Topological phase diagram of SrSi$_2$ is mapped out as a function of the lattice constant with focus on the semimetal order. A tunable double-Weyl Fermion state in Sr$_{1-x}$Ca$_{x}$Si$_2$ and Sr$_{1-x}$Ba$_{x}$Si$_2$ alloys is identified. Ca doping in SrSi$_2$ is shown to yield a double-Weyl semimetal with a large Fermi arc length, while Ba doping leads to a transition from the topological semimetal to a gapped insulator state. Our study indicates that SrSi$_2$ materials family could provide an interesting platform for accessing the unique topological properties of Weyl semimetals.

\end{abstract}

\maketitle

\section{Introduction}
\begin{figure}[t!] %Bulk crystal structure
\includegraphics[width=0.8\textwidth]{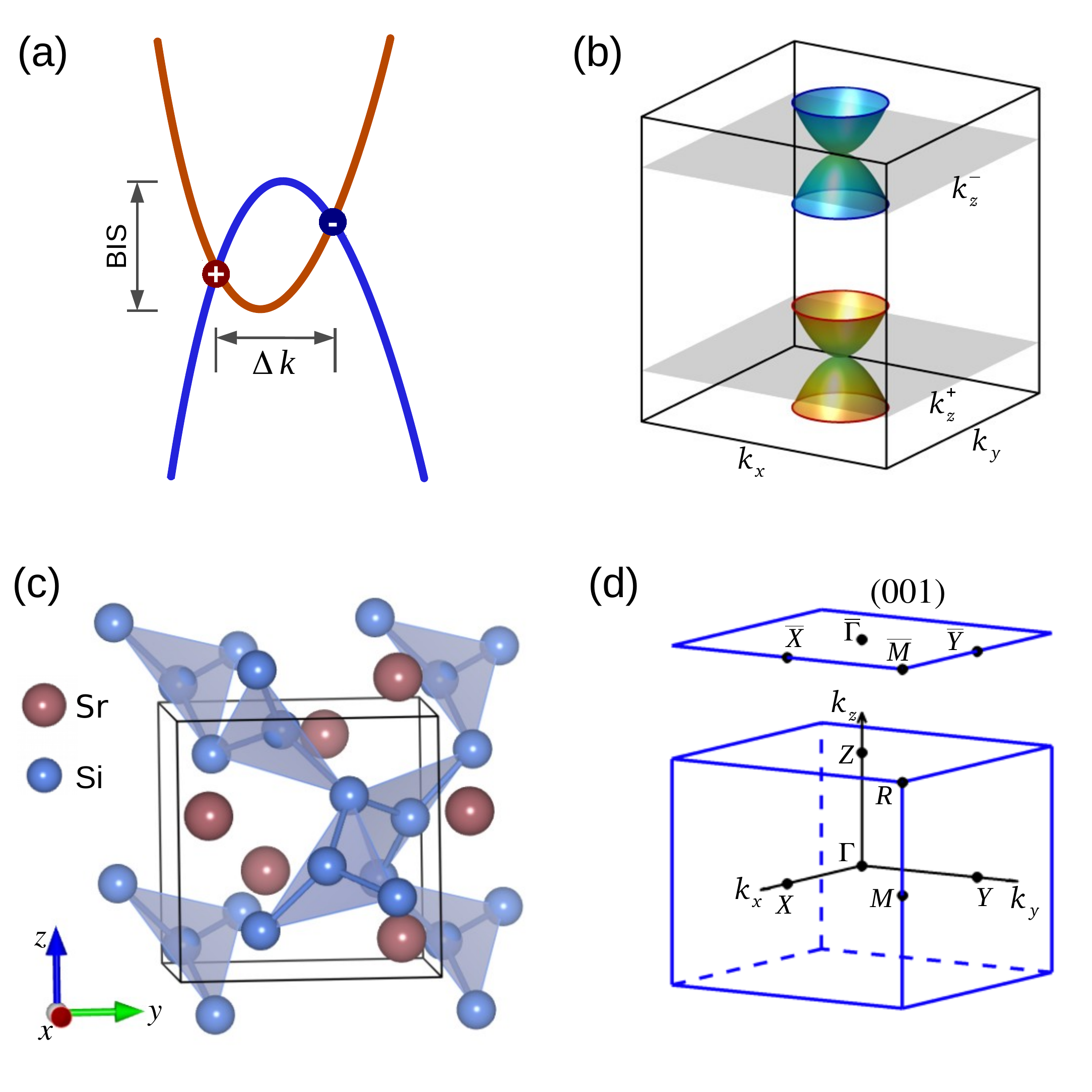} 
\caption {{\bf Double-Weyl fermions and crystal structure of SrSi$_2$.} (a) Schematic of the pair of band-inversion Weyl nodes (gapless bulk crossing points) in a crystal without mirror symmetry. Red (positive) and blue (negative) filled circles mark Weyl nodes of opposite chirality. The topological strength of Weyl semimetal states is determined by the band inversion strength (BIS) and correlates with the separation between the Weyl fermions ($\Delta k$) of opposite chiral charge. (b) The quadratic band dispersion of a pair of double-Weyl fermions (chiral charge $\pm 2$) at $k_z=+k_z^{\pm}$ planes in the BZ. Superscripts $\pm$ denote the chirality of double-Weyl fermions. (c) Bulk chiral crystal structure of SrSi$_2$ with a silicon network with coordination of 3. (d) The primitive bulk BZ for the cubic unit cell with four inequivalent high symmetry $k$-points, $\Gamma(0,0,0)$, $X(\pi,0,0)$, $M(\pi,\pi,0)$, and $R(\pi,\pi,\pi)$. Surface (001) plane projected BZ with high-symmetry points is also shown.} \label{fig:bulkCS}
\end{figure}

Recent discovery of three-dimensional (3D) Weyl semimetals (WSMs) has opened a new research direction in condensed matter physics and materials science with tremendous potential for conceptual novelties and device applications\cite{HWeyl_theory1929,FDNielsen1981,alderbellNenomiya1983,weyl_balents2011,Report_Jia2016,ReviewAshvin2013,ReviewBansil2016,Review_hasan2015}. In WSMs, the bulk system hosts gapless crossing points, called Weyl nodes, between the valence and conduction bands through the time reversal or inversion symmetry breaking mechanism\cite{weyl_TPTsuichi2007,weyl_IASBalents2012,weyl_IASTRBBurkov2012}. Each Weyl node carries a definite chiral charge and acts as a monopole of the Berry curvature in momentum space. Since the total chiral charge in the bulk Brillouin zone (BZ) must be zero, the Weyl nodes always come in pairs\cite{FDNielsen1981,alderbellNenomiya1983}. The two Weyl nodes in each pair are separated in momentum space but carry equal and opposite chiral charges. The Weyl nodes are robust against perturbations that preserve translation invariance and can only be annihilated in pairs of opposite chirality. The $k$-space separation of Weyl nodes can be taken as a measure of the topological strength of a Weyl semimetal state (see Fig. \ref{fig:bulkCS}(a)). A WSM exhibits many fascinating properties such as open Fermi surfaces (Fermi arcs) in its surface electronic structure, negative magnetoresistance, the chiral anomaly, and possibilities for novel superconductivity and next-generation electronics/spintronics devices\cite{weyl_TPTsuichi2007,weyl_IASBalents2012,weyl_IASTRBBurkov2012,fermiarcs_Xu2015,chiralmagnetic2008,chiralanomaly_Burkov2012,QHE_weylsemimetal,weyltrans1_Hosur2013,superC_hosur2014,magnetopt_Ashby2013}. 

Based on the energy dispersion around a Weyl node, WSMs are classified into two types. A Type-I WSM displays a conical energy-dispersion, so that the associated Fermi surface shrinks to a single discrete point with a vanishing density of states (DOS) at node energy. In contrast, a Type-II WSM exhibits a strongly tilted-over cone so that the Weyl node is formed at the boundary between the electron and hole pockets with a finite DOS at the node energy. WSMs have been proposed in many candidate compounds and  alloys\cite{weyl_ashwin2011,double_HgCr2Se4,TaAs_theorySM,TaAs_theoryWeng,WTe2_Soluyanov2016,MoTe2_SunYan2015,MWTe2_Tayrong2016,RAlGe_theory,alloy_singh2012,alloy_HgMnTe2014,alloy_LaBaTe2014,SeTe_motoaki2015,multi_Burkov2011,Ta3S2_Guoqing,TaIrTe4_theory}, and a few of these materials have been realized experimentally through the observation of Weyl cones and Fermi arc surface states\cite{TaAs_expXu2015,TaAs_expLv2015,TaP_expXu2015,NaAs_expXu2016,LaAlGe_exp,MWTe2_Ilya2016,MoTe2_exp,TaIrTe4_expHau,TaIrTe4_expIlya}. 
In particular, Type-I WSMs have been demonstrated in TaAs materials class\cite{TaAs_theorySM,TaAs_theoryWeng,TaAs_expXu2015,TaAs_expLv2015,TaP_expXu2015,NaAs_expXu2016} whereas Type-II WSMs have been realized in LaAlGe\cite{LaAlGe_exp}, Mo$_x$W$_{1-x}$Te$_2$\cite{MWTe2_Ilya2016,MoTe2_exp}, and TaIrTe$_4$\cite{TaIrTe4_expHau,TaIrTe4_expIlya}.

The transitional-metal monophosphides (TaAs, TaP, NbAs, and NbP) were the first WSMs to be realized experimentally, and have been explored quite extensively in connection with their unique topological states and transport characteristics\cite{TaAs_theorySM,TaAs_theoryWeng,TaAs_expXu2015,TaAs_expLv2015,TaP_expXu2015,NaAs_expXu2016}. These materials exist in stoichiometric single-crystalline phase, and host a robust WSM state through the breaking of inversion symmetry. They carry a total of 12 pairs of single-Weyl nodes (chiral charge of $\pm 1$) with an energy dispersion that is linear in all three dimensions in the bulk BZ. Beyond such single-Weyl semimetals, the existence of a new type of 3D topological semimetal with a higher chiral charge of $\pm2$, dubbed a double-Weyl semimetal, has been proposed where the protection comes from the $C_4$ or $C_6$ rotational symmetry \cite{double_HgCr2Se4,double_multiRot,double_SrSi2SM,Multi_Hei,doubleW_correl,doubleW_correlMott}. In effect, these point-group symmetries bring two single-Weyl nodes together into a higher-symmetry point/axis, and result in the double-Weyl node around which the energy dispersion is quadratic along two momentum dimensions and linear along the third dimension, see Fig. \ref{fig:bulkCS}(b).

The double-WSM state has been predicted recently in the inversion-asymmetric chiral compound strontium disilicide, SrSi$_2$, through band structure calculations\cite{double_SrSi2SM}. This is an interesting material because it is composed of non-toxic and naturally abundant elements, and it has been known for decades as a promising candidate for thermoelectric applications through chemical substitution \cite{SrSi2_cyststr,SrSi2_alloys,SrSi2_ThermoEalloy1,SrSi2_ThermoEalloy2,SrSi2_Thermo3,SrSi2_transport}. The double-Weyl nodes in SrSi$_2$ are generated via a band inversion mechanism, and lie along the $C_4$ rotation axis as seen in Figs. \ref{fig:bulkCS}(a) and (b). The $k$-space separation of pairs of Weyl nodes can be related to the band-inversion-strength (BIS), which can thus be tuned by pushing the valence and conduction bands apart, while all the symmetries involved are preserved. An estimate of BIS requires an accurate determination of the orbital occupations and the associate band edges. The experimental situation with SrSi$_2$, however, remain uncertain. Early transport measurements\cite{SrSi2_transport} reported SrSi$_2$ to be a narrow band gap ($10 \sim 30$ meV) semiconductor, while more recent experiments find it to be a gapless semimetal\cite{SrSi2_ThermoEalloy1,SrSi2_ThermoEalloy2}. Other experimental studies focusing on thermoelectric properties show that the semimetal state of SrSi$_2$ could be stabilized through chemical substitution of Sr by the lighter Ca atoms \cite{SrSi2_ThermoEalloy1,SrSi2_ThermoEalloy2}.

Here, we address the electronic ground state and topological properties of SrSi$_2$ using {\it ab-initio} calculations with the Heyd-Scuseria-Ernzerof (HSE) exchange-correlation functional\cite{HSE_Perdew1996}. This hybrid functional incorporates a part of the exact Fock exchange and it is known to yield improved results. Our analysis reveals that even though the double-WSM state in SrSi$_2$ is robust, the overlap between the valence and conduction bands is small, and for this reason SrSi$_2$ lies close to the phase boundary between a topological semimetal and a fully gapped state. We present a phase diagram of topological order in the lattice parameter space, and predict that SrSi$_2$ could realize a tunable double-WSM state through alloying in Sr$_{1-x}$Ca$_{x}$Si$_2$ and Sr$_{1-x}$Ba$_{x}$Si$_2$. A double-WSM with increased topological strength and a topological metal to insulator transition are thus possible with Ca and Ba substitution of Sr in SrSi$_2$. Notably, a tunable single-Weyl semimetal state has been predicted in Mo$_x$W$_{1-x}$Te$_2$ \cite{MWTe2_Tayrong2016}.

\section{Methodology and structural properties}

Electronic structures were calculated with the projector augmented wave (PAW) method\cite{vasp,paw} within the density functional theory (DFT)\cite{kohan_dft} framework, using the VASP (Vienna Ab Initio Simulation Package) suite of codes\cite{vasp}. We used both the generalized-gradient-approximation (GGA)\cite{pbe} and the more advanced HSE hybrid-functional\cite{HSE_Perdew1996} to model exchange-correlation effects. Spin-orbit coupling (SOC) was included self-consistently. A plane-wave cutoff energy of 380 eV, and a $\Gamma$-centered $9\times9\times9$ $k$-mesh were used. All calculations were performed by employing the relaxed lattice parameters ($a_{\text{SrSi}_2}$~=~5.564 \r{A}, $a_{\text{CaSi}_2}$~=~5.378 \r{A}, and $a_{\text{BaSi}_2}$~=~6.769 \r{A}) with optimized ionic positions. In order to compute HSE band structures and surface states, we obtained a tight-binding model Hamiltonian by projecting first-principles results onto Wannier orbitals using the VASP2WANNIER90 interface\cite{wannier90,vasp}. The Sr (Ba or Ca) $s$ and $d$ orbitals and Si $s$ and $p$ orbitals were included in the construction of the Wannier functions. Surface state energy dispersions were calculated with the Wanniertools software package\cite{WT_code}.

SrSi$_2$ crystallizes in a simple cubic Bravais lattice with the chiral space group P4$_3$32 ($\# 212$)\cite{double_SrSi2SM,SrSi2_cyststr,SrSi2_alloys}. The unit cell contains four strontium (Sr) atoms and eight silicon (Si) atoms, which occupy the Wyckoff positions 4a and 8c, respectively. The Si atoms form a three-dimensional network in which each silicon atom is connected to its three neighboring atoms, see Fig. \ref{fig:bulkCS}(c). Due to this unique crystal structure, SrSi$_2$ lacks both mirror and inversion symmetries. Nevertheless, the C$_4$ rotation symmetry is preserved with the rotation axis lying along the three principal axes. Figure \ref{fig:bulkCS}(d) shows the bulk BZ with the four inequivalent high-symmetry points $\Gamma$, $X$, $M$, and $R$ marked. Note that $X$, $Y$, and $Z$ are equivalent points that lie on the corresponding $C_4$ rotation axis ($k_x$, $k_y$ and $k_z$).

\section{Results and Discussion}
\subsection{Bulk topological structure}

\begin{figure}[ht!] % SrSi2 bulk band structure
\includegraphics[width=0.9\textwidth]{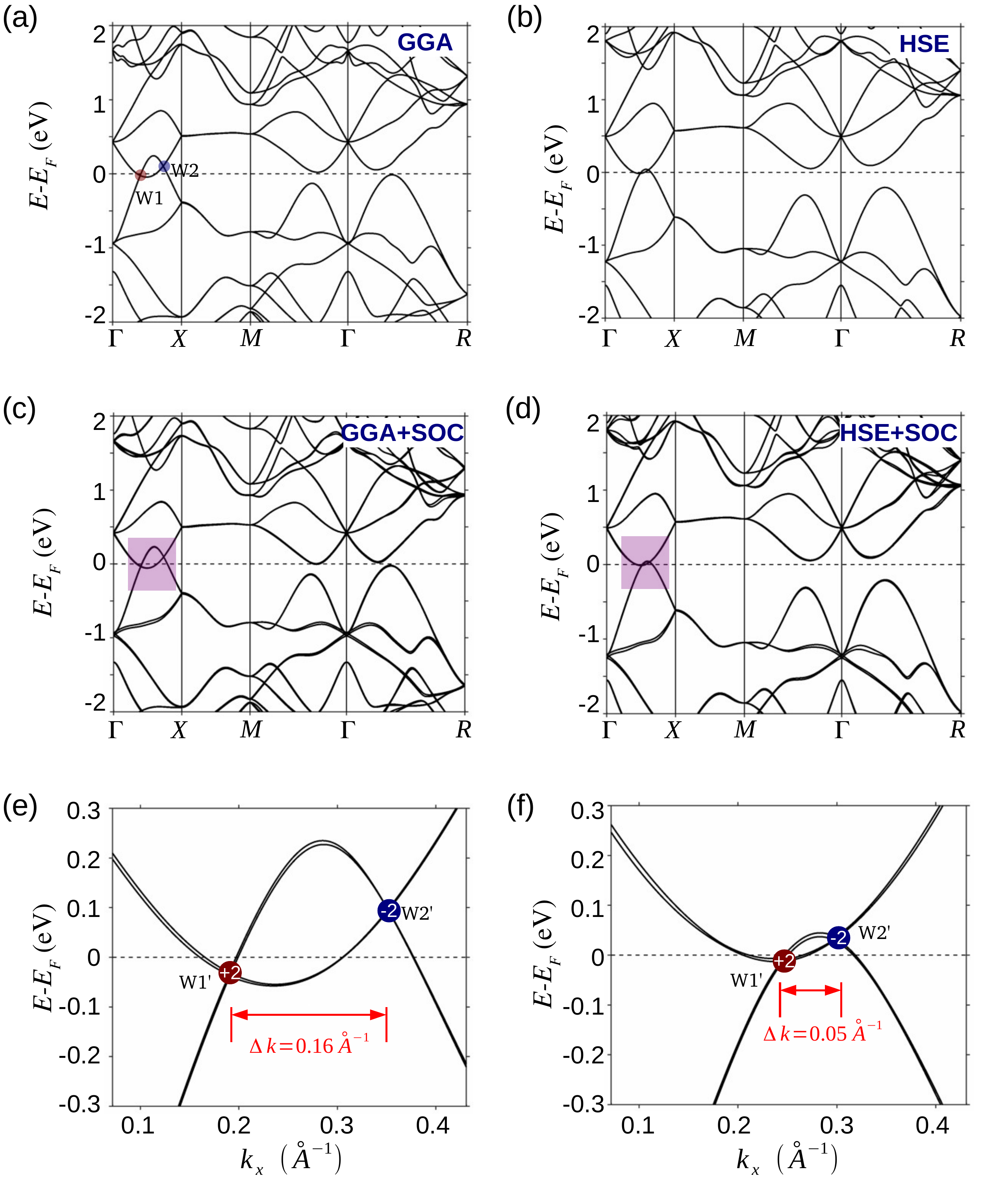} 
\caption{{\bf Topological bulk electronic structure of SrSi$_2$}. Bulk band structure calculated with GGA [(a) and (c)] and HSE [(b) and (d)] functionals. Top and middle rows show band structures without and with SOC, respectively. Dashed horizontal lines mark the Fermi level. (e) and (f) Closeup of the area highlighted by violet boxes in (c) and (d). Red and blue markers identify the double-Weyl nodes of opposite chiral charge. The BIS and separation between the Weyl nodes, $\Delta k$, reduce considerably in band structures obtained with the HSE compared to the GGA functional.}
\label{fig:bulkBS}
\end{figure}

The bulk electronic structure of SrSi$_2$ calculated using the GGA without the SOC is shown in Fig. \ref{fig:bulkBS}(a). The valence and conduction bands clearly dip into each other along the $\Gamma-X$ line, suggesting a semimetallic state with an inverted band structure. Consistent with other recent band structure calculations, the valence and conduction bands cross at two discrete $k$-points on the $\Gamma-X$ line, labeled W1 and W2 in Fig. \ref{fig:bulkBS}(a). Since the $\Gamma-X$ line is a $C_4$ rotation axis of the cubic lattice, all bands on this axis have well defined $C_4$ rotational eigenvalues. From our first-principles Bloch wave functions, we find that the valence and conduction bands in the vicinity of W1 have rotational eigenvalues of $i$ and $-1$, respectively. By applying the recently proposed method for determining the chiral charge of a Weyl node lying on a rotation axis\cite{double_multiRot}, we find that W1 has a nonzero chiral charge of $+1$, while W2 has an equal and opposite charge of $-1$. We have also analyzed the chiral charge by calculating the Berry flux on a closed surface enclosing the Weyl nodes to ascertain that SrSi$_2$ is a Weyl semimetal without the SOC, consistent with the results of Ref. ~\cite{double_SrSi2SM}. Interestingly, a pair of W1 and W2 Weyl nodes locates at different energies. This is because of the chiral crystal structure of SrSi$_2$.

When the SOC is included, bands with the same rotational eigenvalue hybridize while those with opposite eigenvalues remain gapless as shown in Fig. \ref{fig:bulkBS}(c); the gapless points are denoted by W1' and W2' in the closeup of Fig. \ref{fig:bulkBS}(e). By analyzing eigenvalues and appropriate integrals of Berry flux, we find that W1' and W2' carry equal and opposite higher chiral charges of $\pm 2$ as marked in Fig. \ref{fig:bulkBS}(e). The energy dispersion around these Weyl nodes is quadratic along two dimensions and linear along the third dimension as shown schematically in Fig. \ref{fig:bulkCS}(b). These results clearly show that SrSi$_2$ is a double-WSM. This should be contrasted with the case of materials such as Ta$_3$S$_2$ where each spinless Weyl node in the absence of SOC evolves into two spinful Weyl nodes when the SOC is included\cite{Ta3S2_Guoqing}.

The bulk band structures based on the HSE functional are presented in Figs. \ref{fig:bulkBS}(b,d) without and with the inclusion of SOC, respectively. These results show two Weyl-node crossings on the $\Gamma-X$ line, similar to the GGA results. However, the valence and conduction bands are now pushed in opposite directions such that the energy range over which they overlap is substantially reduced, but the double-WSM state is till seen to survive. In order to quantify the semimetallic character, we define the BIS as the difference between the minimum of the conduction bands and the maximum of the valence bands along the $\Gamma-X$ direction [Fig. \ref{fig:bulkCS}(a)]. [A negative value of BIS indicates a semimetal while a positive value identifies an insulator.] The GGA-based value of the BIS is found to be -0.287 eV, whereas with HSE, it decreases by about 80\% to -0.041 eV, see Figs. \ref{fig:bulkBS}(e,f). Similarly, the momentum space separation, $\Delta k$, between a pair of Weyl nodes reduces from a value of 0.16 \r{A}$^{-1}$ with GGA to 0.05 \r{A}$^{-1}$ with HSE. The small HSE-based value of the BIS or $\Delta k$ indicates that even though the topological semimetal state in SrSi$_2$ remains intact, this material lies close to the phase boundary between a topological semimetal and a fully gapped insulator state, which might explain inconsistent experimental observations in  SrSi$_2$\cite{SrSi2_ThermoEalloy1,SrSi2_ThermoEalloy2}.

\subsection{Surface electronic structure and Fermi arc states}
\begin{figure*}[ht!] % topological surface states
\includegraphics[width=0.9\textwidth]{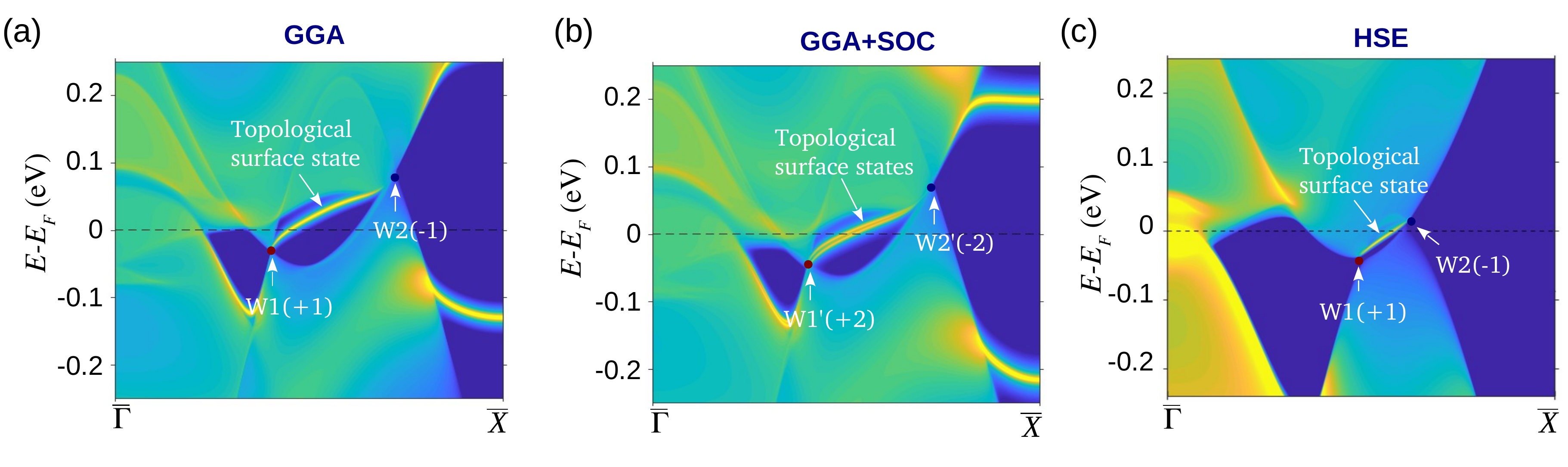} 
\caption{{\bf Topological surface states of SrSi$_2$.} (a) Surface state spectrum for the (001) surface without SOC using the GGA functional. Sharp yellow lines identify surface states. The chiral topological surface states emanate from and terminate at the surface projected Weyl nodes W1(+1) and W2(-1). (b) Same as (a) but with the inclusion of SOC. The double chiral topological states connect Weyl nodes of opposite chirality, which are consistent with the calculated chiral charge of $\pm2$. (d) Surface state spectrum of SrSi$_2$ obtained with HSE functional without SOC. Length of the Fermi arc states is reduced compared to (a), which reflects the decreased topological strength.}
\label{fig:FermiSS}
\end{figure*}

In Fig. \ref{fig:FermiSS}, we present the surface electronic structure of (001) surface of SrSi$_2$ obtained with different energy functionals. The unique signature of a WSM state is the existence of Fermi arc states that terminate at the projections of the bulk Weyl nodes. In the existing WSMs such as TaAs, a pair of Weyl nodes separated in momentum space is located at the same energy. By contrast, in SrSi$_2$, as a result of the chiral crystal structure without any mirror symmetry, the Weyl nodes in a pair are located at different energies. This makes it difficult to visualize the connection of the Fermi arcs with a pair of Weyl nodes via a constant energy cut over the surface\cite{double_SrSi2SM}. For this reason, we present the surface band structure in the projected 2D BZ in discussing the topological surface states below.

Figure \ref{fig:FermiSS}(a) shows the energy dispersion along the $\overline{\Gamma}-\overline{X}$ surface BZ line (Fig. \ref{fig:bulkCS}(d)) that passes through a pair of bulk Weyl nodes, W1(+1) and W2(-1) where +1 and -1 denote positive and negative unit chiral charge, respectively. A single topological surface state is seen clearly that directly connects a pair of projected Weyl nodes. This indicates that the Chern number associated with the 2D slice passing between the Weyl nodes is 1 in accord with the calculated Weyl node chiral charge from Berry flux of $\pm 1$ without the SOC. As shown in Fig. \ref{fig:FermiSS} (b), in the presence of the SOC, there are two topological chiral surface states that connect the projected Weyl nodes W1'(+2) and W2'(-2). The double-chiral states carry the Chern number 2 in accord with the higher chiral charge of $\pm$2. Note that these states terminate directly at the projected Weyl nodes as expected. In the HSE-based band structure in Fig. \ref{fig:FermiSS} (c) also these surface states continue to reside on the $\overline{\Gamma}-\overline{X}$ line. However, the length of the related Fermi arc shrinks substantially, reflecting the smaller separation between the Weyl nodes, and push the material closer to a topological critical point.

\subsection{Tunable double-Weyl state and topological phase transition}

\begin{figure}[ht!] %topological phase diagram
\includegraphics[width=0.8\textwidth]{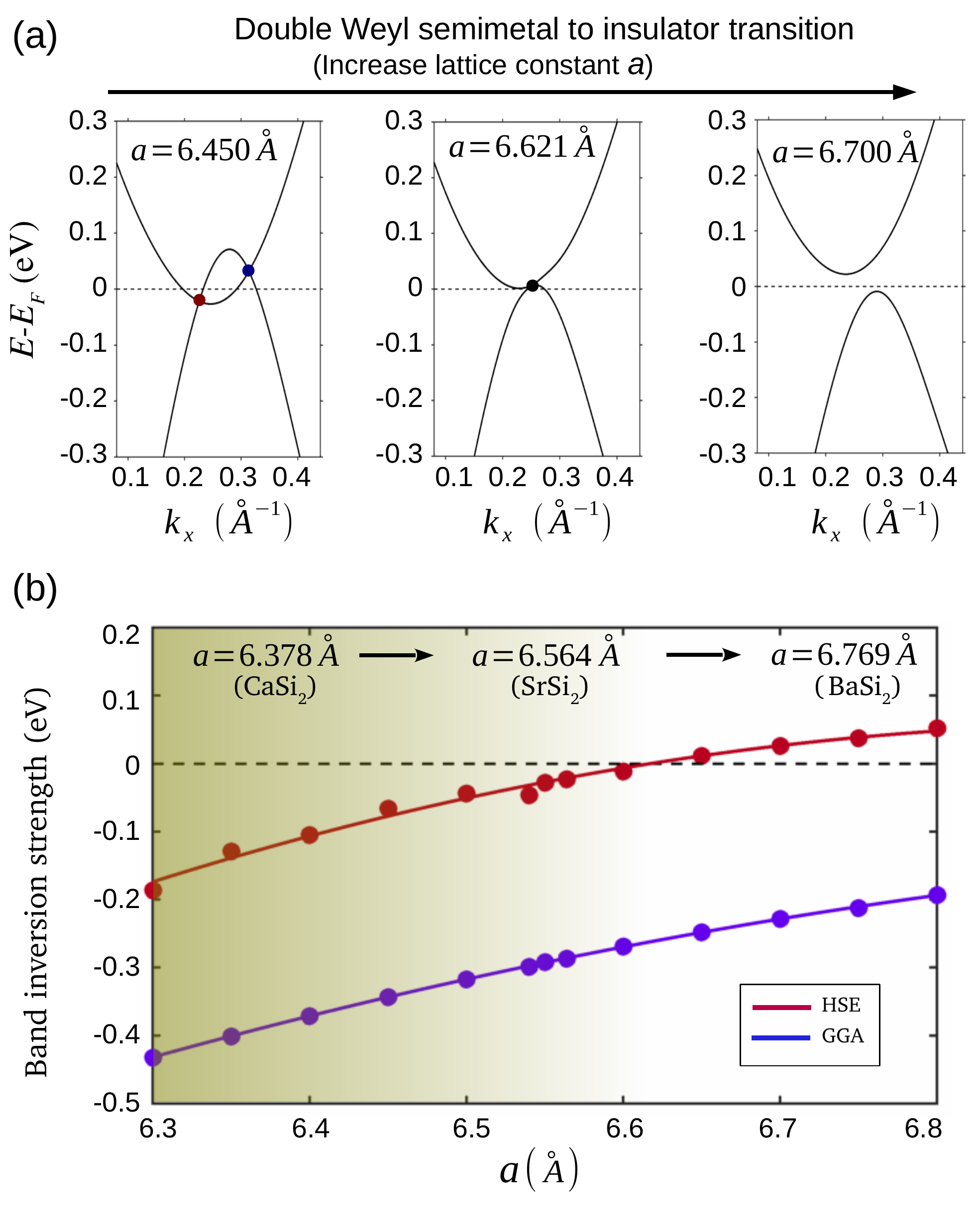} 
\caption{ {\bf Topological semimetal-to-insulator transition in SrSi$_2$.} (a) Energy dispersions along $k_x$ direction for an irreducible pair of Weyl nodes as the lattice constant $a$ is varied. At $a$ = 6.450 \r{A} (left panel), we observe two Weyl nodes with a quadratic energy dispersion and chiral charge $\pm2$. The momentum space separation of Weyl nodes and the BIS reduce considerably for a = 6.621 \r{A} where a pair of Weyl nodes annihilate with each other (middle panel). A gap emerges for  $a$ = 6.700 \r{A} (right panel). (b) The BIS for a range of lattice constant values of SrSi$_2$. The red and violet circles identify BIS computed with the HSE and GGA functionals, respectively. Red and violet lines are guide to the eye. The dashed zero line corresponds to the topological critical point between the double-WSM (BIS $ < 0$) and fully gapped state (BIS $ > 0$).}  
\label{fig:bulkPD}
\end{figure}

We discuss the topological double WSM-to-insulator transition in SrSi$_2$ with reference to Fig. \ref{fig:bulkPD}.  For this purpose, it is helpful to see energy dispersion along the $\Gamma-X$ direction on which an irreducible pair of double Weyl nodes is located; Fig. \ref{fig:bulkPD}(a) shows the dispersion for a series of values of the lattice constant $a$. The bands are seen in Fig. \ref{fig:bulkBS}(f) to cross at discrete points to form a pair of Weyl nodes corresponding to the original value of the lattice constant. When the lattice constant is decreased by 1.7 \% to $a=6.450 \text{\r{A}}$, there is an increase in the Weyl node separation or equivalently the BIS. However, when the lattice constant is increased by just 0.87 \% to $a=6.621 \text{\r{A}}$, the separation between the Weyl nodes vanishes and the system reaches a topological critical point. With further increase in lattice constant to $a=6.700 \text{\r{A}}$, the two Weyl nodes annihilate and the energy spectrum becomes fully gapped. We checked the gapped state for possible topological order, but it is found to be topologically trivial. Therefore, by increasing the lattice constant, one can go through a topological phase transition from a double-WSM state to a trivial insulator. Evolution of the BIS parameter over a wide range of values of the lattice constant is shown in Fig. \ref{fig:bulkPD}(b). Here, The negative values of BIS refer to a double-WSM state and positive value to the gapped insulator [red curve in Fig. \ref{fig:bulkPD}(b)]. Notably, the GGA-based BIS values remain negative over a large range of $a$. 

The aforementioned topological phase transition should be amenable to access in experiments on SrSi$_2$. \cite{SrSi2_alloys,SrSi2_ThermoEalloy1,SrSi2_ThermoEalloy2,SrSi2_transport,SrSi2_Thermo3} Based on the dependence of free energy on the lattice constant, we estimate that a negative pressure of $\sim$1.5 GPa would be sufficient to induce this transition\cite{murnaghan}. Such a pressure could be realized, for example, by partial substitution of Sr by Ba in Sr$_{1-x}$Ba$_{x}$Si$_2$ alloys. Since Ba has a larger atomic size, the replacement of Sr by Ba causes the lattice to expand, or equivalently, to produce a negative chemical pressure\cite{SrSi2_alloys,SrSi2_ThermoEalloy1,SrSi2_ThermoEalloy2}. Note that Ba-doped SrSi$_2$ has already been explored and found to stabilize in an insulator state as reported in transport experiments, and further discussed below.  

\begin{figure}[ht!]% Ca(Ba)Si2 band structure
\includegraphics[width=0.9\textwidth]{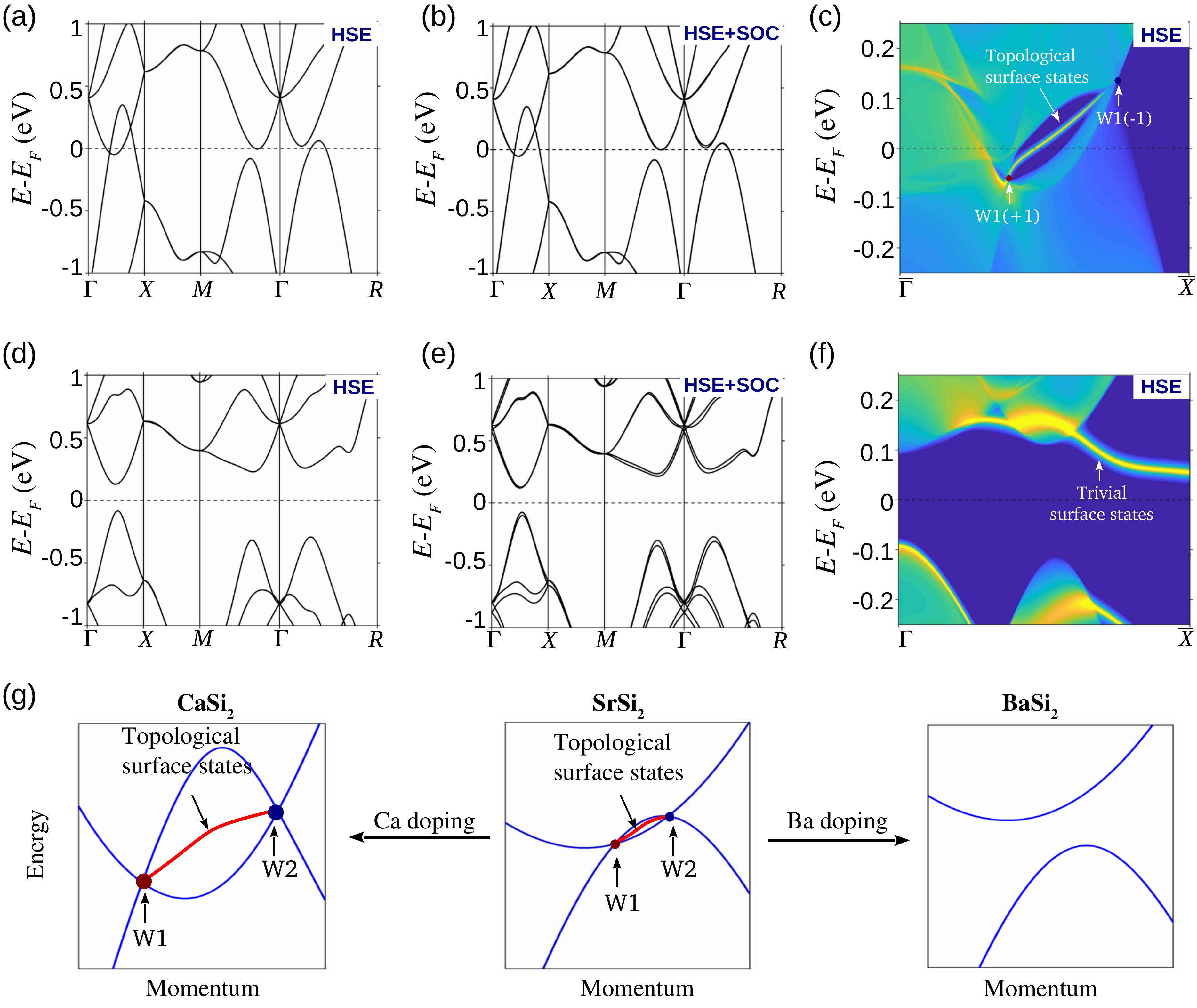} 
\caption{{\bf Tunable topological semimetal state in Ca$_{1-x}$Sr$_x$Si$_2$ and Ba$_{1-x}$Sr$_x$Si$_2$.} (a) and (b) HSE-based bulk band structure of CaSi$_2$ without and with SOC, respectively. We observe two single-Weyl nodes without SOC and double Weyl nodes with SOC on the $\Gamma-X$ line. (c) Surface electronic spectrum for (001) surface of CaSi$_2$. The chiral topological surface states over an extended $k$ space region show an increased topological strength of double-Weyl semimetal state in CaSi$_2$. (d)-(f) Same as (a)-(c) but for BaSi$_2$. The absence of bulk gapless crossings and non-trivial surface state confirm its insulating character. (g) Schematic band diagram of CaSi$_2$ (left panel), SrSi$_2$ (middle panel), and BaSi$_2$ (right panel). Evolution of the system as a function of Ca or Ba substitution in SrSi$_2$ is shown with arrows. Topological metallic strength increases with Ca substitution whereas Ba doping leads to a fully gapped insulator state. } 
\label{fig:BaCaSi2}
\end{figure}

We now turn to consider the tunability of the double-WSM state in SrSi$_2$ based material class with reference to Fig. \ref{fig:BaCaSi2}. It has been known for some time that Ca and Ba substitution of Sr in SrSi$_2$, i.e., Sr$_{1-x}$Ca$_{x}$Si$_2$ and Sr$_{1-x}$Ba$_{x}$Si$_2$ enhances room-temperature thermoelectric properties\cite{SrSi2_alloys,SrSi2_ThermoEalloy1,SrSi2_ThermoEalloy2,SrSi2_transport,SrSi2_Thermo3}. Also, Ca substituted Sr$_{1-x}$Ca$_{x}$Si$_2$ alloys have been found to be more metallic than SrSi$_2$, while Ba substitution Sr$_{1-x}$Ba$_{x}$Si$_2$ alloys results in a semiconducting behavior \cite{SrSi2_alloys,SrSi2_ThermoEalloy1,SrSi2_ThermoEalloy2,SrSi2_transport,SrSi2_Thermo3}. Keep in mind these results and our discussion above about the phase transition in SrSi$_2$ as a function of the lattice constant, and that Ca-doping will shrink the lattice while Ba doping will expand the lattice. It follows then that the end compounds CaSi$_2$ and BaSi$_2$ lie electronically on the opposite sides of SrSi$_2$. In other words, Ca-doping in SrSi$_2$ will be expected to increase its topological strength whereas Ba-doping reduces it and pushes the material toward the trivial insulator state. 

In Fig. \ref{fig:BaCaSi2}, we present the bulk and surface energy spectra of the end compounds CaSi$_2$ and BaSi$_2$. The HSE-based bulk band structure of CaSi$_2$ without and with SOC in Figs. \ref{fig:BaCaSi2}(a,b), respectively, is similar to that of SrSi$_2$. A pair of Weyl nodes is seen on the rotation axis. In comparison to SrSi$_2$, the BIS and momentum space separation of Weyl nodes are enhanced. The surface band structure shows the presence of the topological chiral state that connects the pair of projected Weyl nodes over the surface. In contrast to CaSi$_2$, the bulk band structure of BaSi$_2$ does not show any gapless crossings, indicating its insulating behavior. This is further reflected in the surface state spectrum where we do not see any non-trivial surface state, which connects the projected bulk valence and conduction bands. These results are consistent with our expectation that CaSi$_2$ and BaSi$_2$ lie on the opposite ends of SrSi$_2$ in terms of the BIS strength. Thus the Sr$_{1-x}$Ca$_{x}$Si$_2$ and Sr$_{1-x}$Ba$_{x}$Si$_2$ alloys should allow the realization of tunable double-WSM state by varying the concentration $x$ of Ca or Ba, much like the case of Mo$_x$W$_{1-x}$Te$_2$\cite{MWTe2_Tayrong2016,MWTe2_Ilya2016}.

\section{Conclusion}
In conclusion, we have explored the bulk and surface topological electronic properties of inversion-asymmetric SrSi$_2$ materials class within the framework of the density functional theory framework using the hybrid exchange-correlation functional. SrSi$_2$ is shown to be an exotic semimetal hosting a pair of double-Weyl nodes with a chiral charge of $\pm 2$. The separation between these double-Weyl nodes is significantly smaller than the earlier GGA-based computations in the literature, and we thus adduce that the material lies close to a topological semimetal-to-insulator transition point. In this connection, we consider how topological phases evolve in SrSi$_2$ under tensile and compressive strains around the equilibrium lattice volume. Insight so gained, allows us to predict that the double-WSM state in Sr$_{1-x}$Ca$_{x}$Si$_2$ and Sr$_{1-x}$Ba$_{x}$Si$_2$ alloys can be tuned by varying the concentration of Ca and Ba dopants. Our study suggests that SrSi$_2$ materials family would provide an interesting new platform for accessing many topological properties of WSMs.

\section*{ACKNOWLEDGMENTS}
Work at the ShenZhen university is financially supported by the Shenzhen Peacock Plan (Grant No. 827-000113, KQTD2016053112042971), Science and Technology Planning Project of Guangdong Province (2016B050501005), and the Educational Commission of Guangdong Province (2016KSTCX126). The work at Northeastern University was supported by the US Department of Energy (DOE), Office of Science, Basic Energy Sciences grant number DE-FG02-07ER46352, and benefited from Northeastern University’s Advanced Scientific Computation Center and the National Energy Research Scientific Computing Center through DOE grant number DE-AC02-05CH11231. H.L. acknowledges the Singapore National Research Foundation for the support under NRF Award No. NRF-NRFF2013-03. M.-C.L acknowledges the support from National Research Foundation of Korea (201500000002559). T.-R.C. is supported by the Ministry of Science and Technology under MOST Young Scholar Fellowship: MOST Grant for the Columbus Program NO. 107-2636-M-006 -004-, National Cheng Kung University, Taiwan, and National Center for Theoretical Sciences (NCTS), Taiwan. S.-M.H. acknowledges the support from the Ministry of Science and Technology (MoST) in Taiwan under Grant No. 105-2112-M-110-014-MY3.

\section*{Author Contributions} 

B.S., G.C., T.-R.C., and M.-C.L. performed the first-principles calculations. B.S. and S.-M.H. performed theoretical analysis. B.S., C.S., H.L., and A.B. wrote the manuscript with help from all authors.

\section*{Competing Interests}

The authors declare no competing interests.

%% Figures

\end{document}